# Comments on the paper: Optical reflectance, optical refractive index and optical conductivity measurements of nonlinear optics for *L*-aspartic acid nickel chloride single crystal

Bikshandarkoil R. Srinivasan, Suvidha G. Naik, Kiran T. Dhavskar
Department of Chemistry, Goa University, Goa 403206, INDIA
Email: srini@unigoa.ac.in Telephone: 0091-(0)832-6519316; Fax: 0091-(0)832-2451184

**Abstract**

We argue that the '*L*-aspartic acid nickel chloride' crystal reported by the authors of the title paper (Optics Communications, 291 (2013) 304–308) is actually the well-known diaqua(*L*-aspartato)nickel(II) hydrate crystal.

**Comment**

The authors of the title paper [1] claim to have grown a so called *L*-aspartic acid nickel chloride (LANC) by the slow evaporation of an aqueous solution containing *L*-aspartic acid (*L*-aspH$_2$), KOH and nickel chloride in 1:1:1 mole ratio at room temperature. The title compound is not represented by a proper chemical formula but instead by an unusual name not in accordance with the nomenclature of chemical compounds and abbreviated by a strange code LANC. No results of single crystal X-ray structure determination, chemical or thermal analysis are given in the paper to understand the exact composition of this so called LANC. Based on an infrared spectral study the authors claim to have identified the presence of (NO$_3$)$^-$ from the peaks at about 1310 cm$^{-1}$ and 829 cm$^{-1}$ in the title compound. This claim only creates confusion because no nitrate containing reagent was used for the crystal growth and *L*-aspH$_2$ does not contain any (NO$_3$)$^-$ group. Unit cell parameters (*a*=13.91 Å, *b*=8.67 Å, *c*=13.50 Å) are supposed to have been calculated from a X-ray powder pattern as per the authors claim '*The recorded data was used to calculate lattice parameter using JCPDS from which it was confirmed that the grown crystals belong to monoclinic crystal system with space group P2$_1$/a*'.

Although the above statement indicates that LANC is a known compound listed in the JCPDS database, the dubious nature of the claim can be evidenced from the assigned centrosymmetric space group $P2_1/a$ for a compound which is supposed to contain an optically active *L*-aspartic acid in its structure. It can be readily pointed out that the assignment is incorrect as any space group containing a glide plane is incompatible with the presence of the chiral *L*-aspartic acid and such structures can only be hosted by a space group (Sohncke) that does not possess mirror or inversion symmetry. For example, the compound diaqua(*L*-aspartato)nickel(II) hydrate represented by the formula [Ni(H$_2$O)$_2$(*L*-asp)]·H$_2$O (*L*-asp = *L*-aspartate dianion) [2] crystallizes in the Sohncke space group $P2_12_12_1$ and is isotructural with the corresponding Co(II) and Zn(II) analogues [M(H$_2$O)$_2$(*L*-asp)]·H$_2$O (M= Co, Ni and Zn) [3].

Based on their optical spectrum the authors declared '*The good transmissions of the crystal in the entire visible region suggest its suitability for second harmonic generation devices.*' However this claim is not only questionable but also unacceptable because Ni(II) compounds are known to be coloured and show characteristic spectra. For example [Ni(H$_2$O)$_2$(*L*-asp)]·H$_2$O exhibits three strong (380, 630 and 1035 nm) and one weak band at 750 nm [2]. Finally mention must be made of the name '*L*-aspartic acid nickel chloride' which indicates the presence of *L*-aspartic acid in a Ni(II) compound isolated from an alkaline medium by use of a strong base like KOH. Since the -COOH group of an amino acid gets deprotonated in alkaline medium to give a carboxylate anion, the presence of free amino acid (*L*-aspartic acid) can be ruled out, showing that the so called '*L*-aspartic acid nickel chloride' is a dubious crystal. As the authors performed their crystal growth reaction by using KOH, the formation of an *L*-aspartate compound of nickel can be expected under the reaction conditions. In order to verify this, we have carried out a slow evaporation crystal growth reaction by using (*L*-aspH$_2$), KOH and commercially available nickel chloride hexahydrate in 1:1:1 mole ratio at room temperature (Scheme 1) and recorded the IR spectrum of the turquoise crystals thus obtained. The spectrum is identical (Fig. 1) with that of an authentic sample of [Ni(H$_2$O)$_2$(*L*-asp)]·H$_2$O prepared by the literature method [2], showing that a so called '*L*-aspartic acid nickel chloride' crystal is actually the well-known diaqua(*L*-aspartato)nickel(II) hydrate crystal. In summary, the above mentioned discussions reveal that the title paper is completely erroneous.

**References**

[1] G. Anbazhagan, P.S. Joseph, G. Shankar, Optical reflectance, optical refractive index and optical conductivity measurements of nonlinear optics for *L*-aspartic acid nickel chloride single crystal, Optics Communications 291 (2013) 304–308.

[2] L. Antolini, L. Menabue, G.C. Pellacani, Structural, Spectroscopic, and Magnetic Properties of Diaqua( *L*-aspartato)nickel(II) Hydrate, J. Chem. Soc. Dalton Trans. (1982) 2541-2543.

[3] T. Doyne, R. Pepinsky, 'Cyclic configuration of the aspartate ion in the crystal structure of zinc, cobaltous and nickelous aspartate trihydrate, Acta Cryst. 10 (1957) 438.

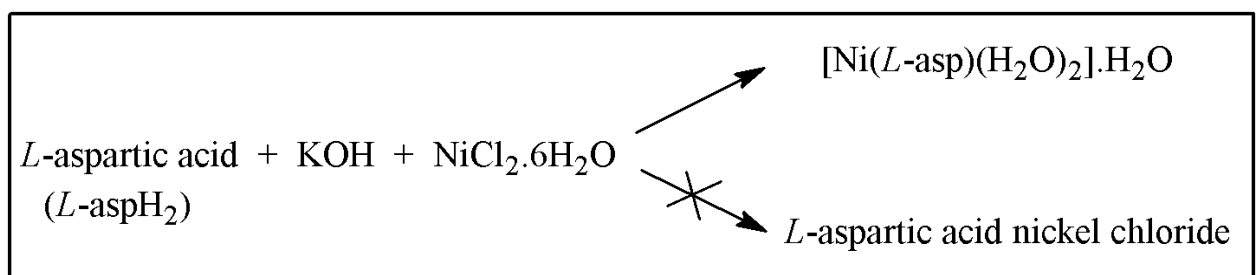

**Scheme 1**

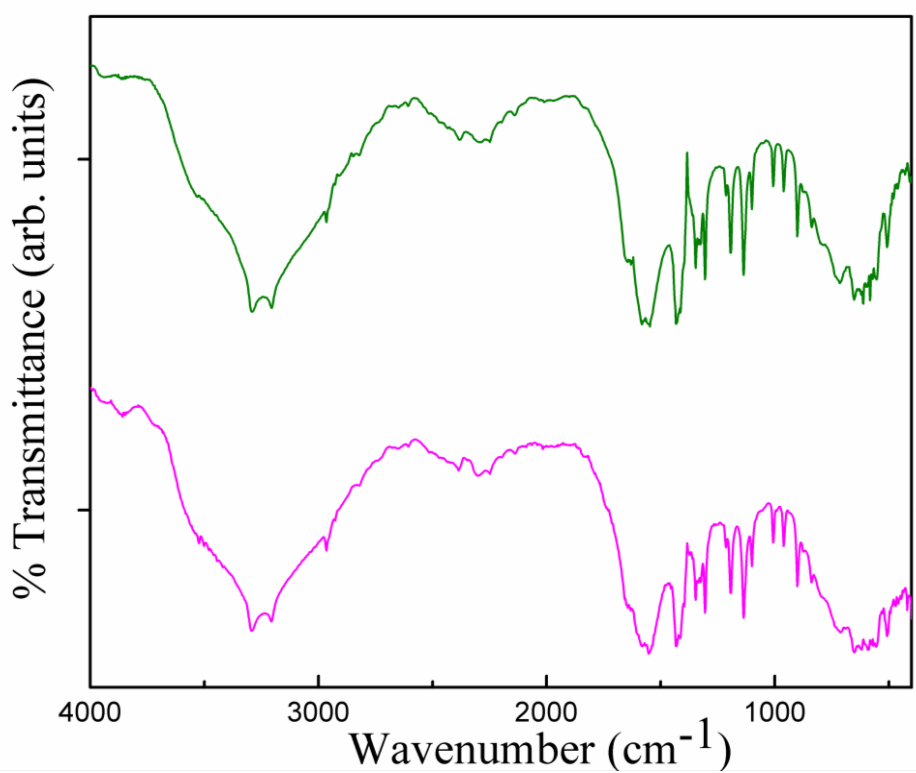

Fig.1: IR spectrum of an authentic [Ni($H_2O$)$_2$(*L*-asp)]·$H_2O$ (top) is identical to that of a so called '*L*-aspartic acid nickel chloride' (bottom) isolated by slow evaporation from an aqueous solution containing (*L*-asp$H_2$), KOH and $NiCl_2$·6$H_2O$.